\documentclass[reprint,
superscriptaddress,
nofootinbib,
amsmath,amssymb,
aps,
]{revtex4-1}

\usepackage{tikz}
\usepackage[utf8]{inputenc}
\usepackage{amsmath,amsfonts,amssymb}
\usepackage{graphicx}
\usepackage[toc,page]{appendix}
\usepackage{cleveref}
\usepackage[loose]{units}
\usepackage{booktabs}
\usepackage{mathtools}
\makeatletter
\gdef\@fpheader{}
\makeatother

\usepackage{graphicx}
\usepackage{dcolumn}
\usepackage{bm}

\begin{document}
	
	\preprint{}
	
	\title{Lattice Simulations of Axion-U(1) Inflation}

	\author{Angelo Caravano}
	\affiliation{	Universit\"ats-Sternwarte M\"unchen, Fakult\"at f\"ur Physik, Ludwig-Maximilians Universit\"at, Scheinerstr. 1, 81679 M\"unchen, Germany}
	\affiliation{Max-Planck-Institut f\"ur Physik (Werner-Heisenberg-Institut),
		F\"ohringer Ring 6, 80805 Munich, Germany}
	
	\author{Eiichiro Komatsu}%
	\affiliation{Max Planck Institute for Astrophysics, Karl Schwarzschild Str. 1, Garching, 85741, Germany}
	\affiliation{Kavli Institute for the Physics and Mathematics of the Universe (Kavli IPMU, WPI), University of Tokyo, Chiba 277-8582, Japan}

	\author{Kaloian D. Lozanov}
	
	\affiliation{
		Illinois Center for Advanced Studies of the Universe \& Department of Physics, University of Illinois at Urbana-Champaign, Urbana, IL 61801, USA.
	}%
	
	\author{Jochen Weller}
	\affiliation{	Universit\"ats-Sternwarte M\"unchen, Fakult\"at f\"ur Physik, Ludwig-Maximilians Universit\"at, Scheinerstr. 1, 81679 M\"unchen, Germany}
	\affiliation{Max Planck Institute for Extraterrestrial Physics,
		Giessenbachstr. 1, 85748 Garching, Germany}

	\begin{abstract}

	
	We present the first nonlinear lattice simulation of an axion field coupled to a U(1) gauge field during inflation. We use it to fully characterize the statistics of the primordial curvature perturbation  $\zeta$. We find high-order statistics to be essential in describing non-Gaussianity of $\zeta$ in the linear regime of the theory. On the contrary, non-Gaussianity is suppressed when the dynamics becomes nonlinear. This relaxes bounds from overproduction of primordial black holes, allowing for an observable gravitational waves signal at pulsar timing array and interferometers scales. Our work establishes lattice simulations as a crucial tool to study the inflationary epoch and its predictions.

		
	\end{abstract}
	
	\maketitle
	\section{Introduction}
	In the standard paradigm, cosmological inflation, the accelerated expansion of the very early universe, is driven by a scalar degree of freedom, the so-called ``inflaton" field $\phi$ \cite{PhysRevD.23.347,Sato:1980yn,Linde:1981mu,PhysRevLett.48.1220,STAROBINSKY198099}. The quantum vacuum fluctuations of $\phi$ provide a natural mechanism to generate the observed anisotropies in the cosmic microwave background (CMB) \cite{Starobinsky:1979ty,Mukhanov:1981xt,HAWKING1982295,PhysRevLett.49.1110,STAROBINSKY1982175,Abbott:1984fp}. Although the simplest single-field slow-roll scenario is compatible with all current observations \cite{Komatsu:2014ioa,Planck:2018jri}, we still lack a complete theoretical understanding of the inflationary universe. For this reason, non-minimal models of inflaton have been studied in the literature, involving multiple scalars or gauge fields. In many interesting cases, the dynamics is affected by nonlinear physics, invalidating the standard perturbative technquiques. Nonlinear lattice simulations, developed to study the reheating epoch after inflation \cite{Khlebnikov_1996,Prokopec_1997,latticeeasy,Frolov_2008,hlattice,Sainio_2012,Child_2013,Easther_2010,Lozanov_2020,figueroa2021cosmolattice}, might be an essential tool to compute predictions from non-minimal inflationary scenarios.
	
	We consider a model where inflation is driven by a pseudoscalar ``axionlike" $\phi$, which is coupled to gauge fields through Chern-Simons interaction $\phi F\tilde F$ \cite{Anber_2006,Anber_2010,Barnaby_2011_Large, Barnaby_2011, Anber_2012,Maleknejad_2011,maleknejad2013gaugeflation,Adshead_2012,Adshead_2013,maleknejad2021su2r}. This system gives rise to unique observational signatures, like non-Gaussianities and chiral gravitational waves, which might be observed with next generation experiments \cite{Komatsu:2022nvu,Campeti:2020xwn}. However, this system is often characterized by strong backreaction effects, associated with a breakdown of perturbation theory \cite{Ferreira:2015omg,Peloso_2016,Papageorgiou:2018rfx,Maleknejad_2019,Lozanov_2019,Mirzagholi:2019jeb,Papageorgiou:2019ecb,Ishiwata:2021yne}. 
	Although significant effort has been put into simulating axion-gauge models during the reheating phase of the universe \cite{Figueroa_2018,Cuissa_2019,Figueroa_2019,figueroa2021cosmolattice,Deskins_2013,Adshead_2015,Adshead_2016,Adshead_2018,Braden_2010,Lozanov_2020,2020axiin,2020axiin2,2021axiin},
	they have never been simulated \textit{during} the inflationary epoch.
	
	In this paper, we use a lattice simulation to study the following axion-gauge system during inflation:
	\begin{align}
		\begin{split}
			S=\int d^4x \sqrt{-g}\Biggl[&\frac{M^2_{\rm Pl}}{2}R-\frac{(\partial_\mu\phi)^2}{2}
			-V(\phi)\\&-\frac{1}{4}F_{\mu\nu}F^{\mu\nu}-\frac{\alpha}{4f}\phi F_{\mu\nu}\tilde{F}^{\mu\nu}\Biggr],
			\label{eq:action}
		\end{split}
	\end{align}
	where $\phi$ is the pseudoscalar inflaton field, coupled to a U(1) gauge field $A_\mu$ with strength tensor $F_{\mu\nu}=\partial_\mu A_\nu - \partial_\nu A_\mu$ and $\tilde F^{\mu\nu}=(2\sqrt{-g})^{-1}\epsilon^{\mu\nu\mu'\nu'}F_{\mu'\nu'}$. Here, $\epsilon^{\mu\nu\mu'\nu'}$ is a totally antisymmetric symbol with $\epsilon^{0123}=1$, $g$ the determinant of the metric tensor, and $R$ the Ricci scalar, $\alpha$ the dimensionless coupling constant, and $f$ the axion decay constant. $M_{\rm Pl}$ is the reduced Planck mass, that we set to 1 throughout this paper. 
	
	The  Chern-Simons interaction $\phi F\tilde F$ leads to an abundant production of gauge field particles \cite{Anber_2006, Anber_2010}, which act as a source for inflaton perturbations $\delta\phi$ via the inverse particle decay \cite{Anber_2010, Barnaby_2011_Large, Barnaby_2011, Anber_2012}, hence affecting the statistical properties of the comoving curvature perturbation $\zeta$.
	The power spectrum of $\zeta$ on super-horizon scales $k\ll aH$ has been estimated analytically as \cite{Anber_2010,Barnaby_2011_Large, Barnaby_2011, Anber_2012}:
	\begin{equation}
		\label{eq:ps_th}
		\mathcal{P}_{\zeta}(k)\simeq\mathcal{P}_{\rm vac}+\mathcal{P}_{\rm vac}^2f_2(\xi)e^{4\pi\xi}, \quad \quad\xi =\frac{\alpha \dot{\phi}}{2fH},
	\end{equation}
	where $\mathcal{P}_{\rm vac}= H^4/(2\pi\dot{{\phi}})^2$ is the vacuum contribution and $f_2(\xi)$ is a function that can be found in Ref. \cite{Barnaby_2011}. This result was derived assuming a constant $\xi$. Moreover, statistics of $\zeta$ are highly non-Gaussian, as $F\tilde F$ is bilinear in the gauge field. The bispectrum has also been estimated in the $\xi-$constant approximation \cite{Anber_2010,Barnaby_2011_Large, Barnaby_2011, Anber_2012}. We avoid reporting the lengthy expression for the bispectrum, whose value can be found, for example, in Ref. \cite{Barnaby_2011}. 
	
	These analytical estimates are valid when at least two key assumptions are satisfied. First, the backreaction of $\phi F\tilde F$ on the background inflationary trajectory is small. This translates into the following bound \cite{Anber_2010,Barnaby_2011_Large, Barnaby_2011, Anber_2012}:
	\begin{equation}
		\label{eq:bounds}
		\frac{H^2}{26\pi|\dot{\phi}|}\xi^{-3/2}e^{\pi\xi}\ll 1.
	\end{equation} 
	Second, the power spectrum $\mathcal{P}_\zeta(k)$ remains small (typically smaller than $10^{-1}$), to ensure perturbativity. A violation of these assumptions invalidates the perturbation theory approach and requires nonlinear tools, such as presented in this paper. 
	
	\section{Lattice simulation}
	Our simulation is based on the methodology developed in Ref. \cite{Caravano_2021,caravano2021lattice}, to which we refer for details. We
	discretize the classical equations of motion in real space.
	We choose to work in the Lorenz gauge $\partial^\mu A_\mu=0$, in which the equations read \cite{Adshead_2015}:
\begin{align}
	\label{eq:nonlin}
	\begin{split}
		&		\phi^{\prime\prime}+2H{\phi^\prime}-\partial_j\partial_j\phi+a^2\frac{\partial V}{\partial \phi}=-a^2\frac{\alpha}{4f}F_{\mu\nu}\tilde F^{\mu\nu}, \\
		&{A}^{\prime\prime}_0-	\partial_j\partial_jA_0=\frac{\alpha}{f}\epsilon_{ijk}\partial_k\phi \partial_iA_j,\\
		&{A}^{\prime\prime}_i-\partial_j\partial_jA_i=\frac{\alpha}{f}\epsilon_{i jk}\phi^\prime \partial_jA_k-\frac{\alpha}{f}\epsilon_{i jk}\partial_j\phi (A^{\prime}_k-\partial_kA_0),
	\end{split}
\end{align}
	where $i,j,k\in\{1,2,3\}$ and the prime denotes derivatives with respect to conformal time. The scale factor in \cref{eq:nonlin} is evolved self-consistently with the second Friedmann equation. 
	As commonly done in the literature \cite{Anber_2010,Barnaby_2011_Large, Barnaby_2011,PhysRevD.85.023525, Anber_2012, PhysRevD.87.103506,domcke2020resonant,gorbar2021gaugefield}, we neglect the role of metric perturbations because gravitational interactions are slow-roll suppressed during inflation. Moreover, the field dynamics of models characterized by large non-Gaussianity is expected to be decoupled from the gravitational sector \cite{Leblond_2011}.
	
	To solve this system of equations, we associate field values $\phi_{n_1,n_2,n_3}$ and ${A}_{\mu,n_1,n_2,n_3}$ to the $N^3$ points of a periodic cubic lattice with comoving volume $L^3$. 
	After defining a discretization scheme for the spatial derivatives, \cref{eq:nonlin} constitute a set of second order coupled differential equations that we solve numerically with a Runge-Kutta 4th order integrator. We start the simulation when the lattice box size satisfies $L\lesssim 1/(aH)$, so that the fields are approximately in their Bunch-Davies vacuum state at the beginning of the simulation. The gauge condition $\partial^\mu A_\mu=0$ is only imposed at the initial time. Therefore, we need to check by hand that $\partial^\mu A_\mu$ vanishes with sufficient precision throughout the evolution. We find that the dimensionless gauge constraint  ${\partial^\mu A_\mu}/{\sqrt{\sum_\rho |\partial^\rho A_\rho|^2}}$ is always smaller than $3\times10^{-4}$ for all the simulation runs shown below.
	
	
	\section{Negligible backreaction}
\begin{figure}
		\centering
		\includegraphics[width=6.8cm]{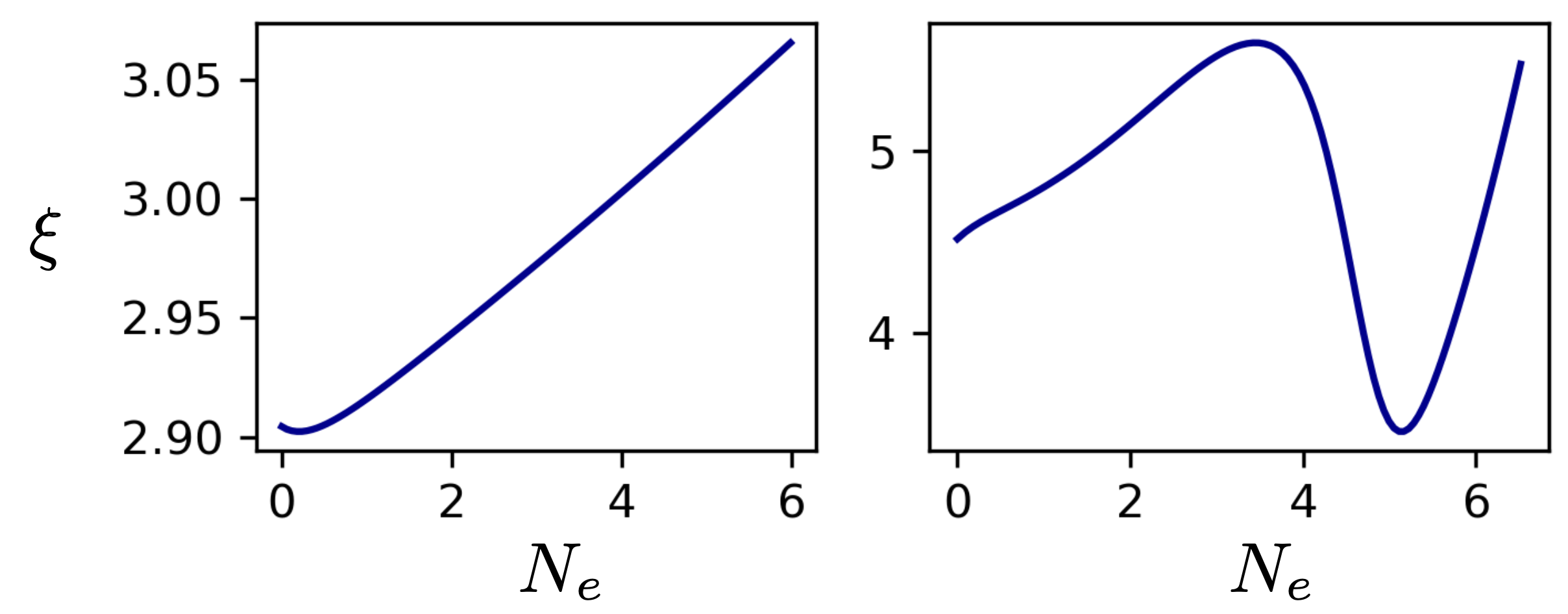}

		\caption{Time evolution of the $\xi$ parameter as a function of the number of e-folds $N_e$, in the case of negligible backreaction (left) and strong backreaction (right).}
		\label{fig:backgroundvalues}
	\end{figure}

	We show the results of the simulation starting from the case when backreaction is negligible (see \cref{eq:bounds}), and compare them to the known analytical results. We assume a monodromy potential for the inflaton \cite{McAllister:2014mpa} $V(\phi)=\frac{1}{2}m^2\phi^2$ with $m=0.51\cdot10^{-5}$. The system is initiated far from the end of inflation by setting ${\phi}=-14.5$. 
	We run a simulation with $N^3=256^3$ points and comoving size $L=2/m$. We evolve the system for $N_e=6$ e-folds, which makes the simulation box satisfy $L\gg 1/(aH)$ at the end of the simulation. For this run, we set the gauge coupling $\alpha/f=42$, which is excluded by CMB observations but allows us to better compare the results of the simulation with the existing analytical estimates. Below we consider a more realistic value of the coupling.
	In the left panel of \cref{fig:backgroundvalues} we show the value of $\xi$ during this simulation, which monotonically grows following the slow-roll trajectory. 
	
\begin{figure}
		\centering
		\includegraphics[width=8cm]{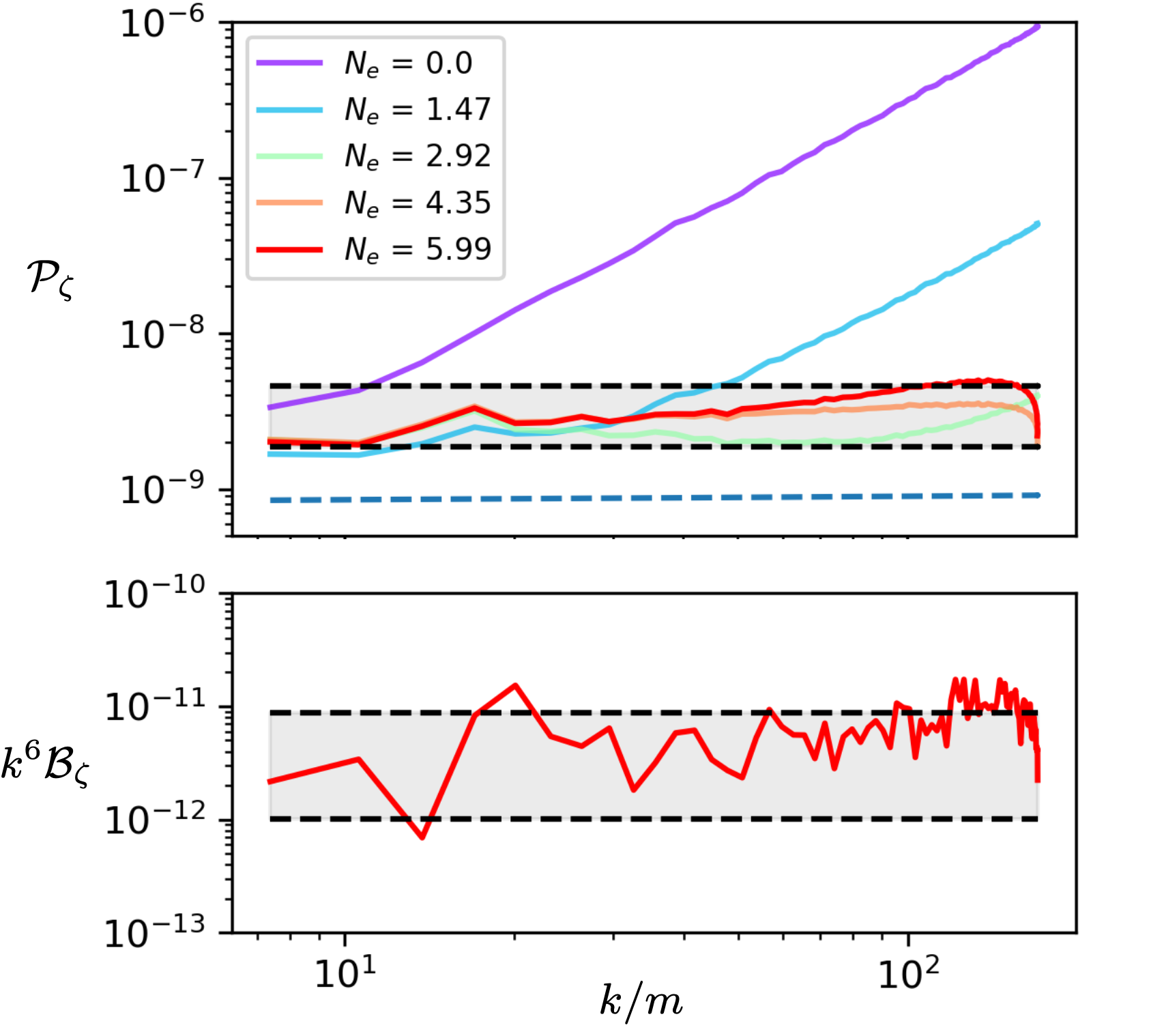}

		\caption{(Top) Power spectrum of $\zeta$ in the case of weak backreaction. The shaded region, delimited by black dashed lines, shows the analytical prediction of \cref{eq:ps_th}. The blue dashed line shows the vacuum contribution $\mathcal{P}_{\rm vac}$. (Bottom) Equilateral bispectrum of $\zeta$ compared to the analytical prediction.}
		\label{fig:ps_weak}
	\end{figure}

	In the upper panel of \cref{fig:ps_weak} we show the power spectrum of the comoving curvature perturbation $\zeta\equiv-\delta\phi H/\dot\phi$ at different times during the simulation\footnote{{This relation for the curvature perturbation is valid as long as the energy density of the Universe is dominated by the background $\phi$ field, which remains true for all the cases considered in this work.}}. We compare the final power spectrum with \cref{eq:ps_th}, which is shown as a shaded region as $\xi$ varies during the evolution. The black dashed lines delimiting this region are computed using the initial and final values of $\xi$. 
	In the bottom panel of \cref{fig:ps_weak} we show the bispectrum $\mathcal{B}_\zeta(k)\equiv\langle\zeta({\vec{k}_1})\zeta({\vec{k}_2})\zeta^{*}({\vec{k}_1}+{\vec{k}_2})\rangle$ on equilateral configurations $k\equiv|\vec{k}_1|=|\vec{k}_2|=|\vec{k}_1+\vec{k}_2|$ at the final time, and compare it to the analytical estimate of Ref. \cite{Barnaby_2011}. 
	We find that both the bispectrum and the power spectrum are in agreement with the analytical estimates.
	Note that for the largest modes there is a drop in the lattice spectra, which is unphysical and it is caused by the lattice UV cutoff. 
	
\begin{figure}
		\centering
		
	\includegraphics[width=8cm]{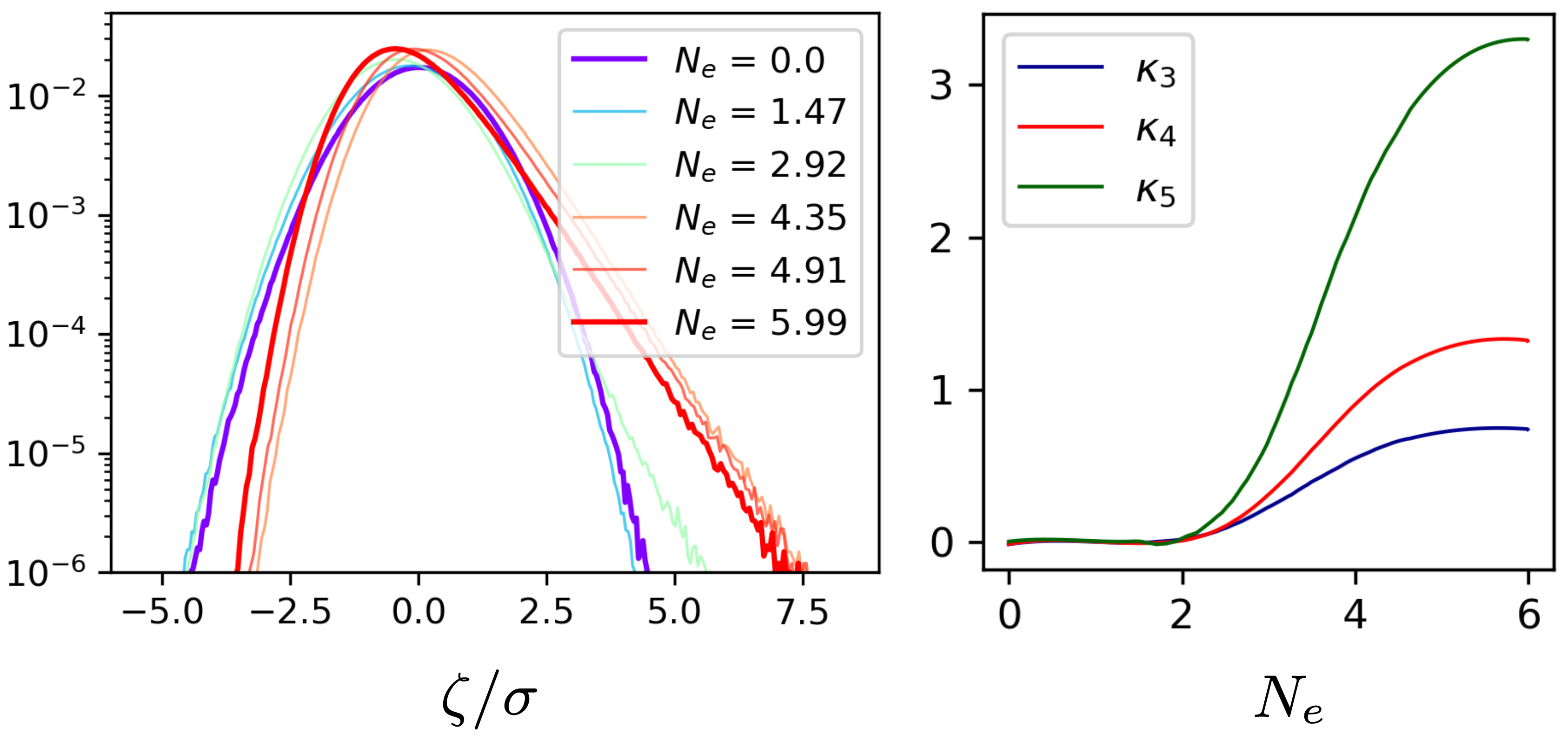}

	\caption{(Left) Normalized histograms of $\zeta$ in real space in the case of weak backreaction. (Right) Time evolution of the correlators defined in \cref{eq:momenta}. For a more detailed version of this figure, with associated errors, see Ref. \cite{Caravano:2022yyv}.}
	\label{fig:hist_weak}
\end{figure}

	Thanks to the lattice approach, we have access to the curvature perturbation in real space. In the left panel of \cref{fig:hist_weak} we show the normalized histograms of the values of $\zeta$ across the $N^3$ points at different times during the simulation.
		We find that non-Gaussianity manifests as a pronounced exponential tail in the distribution of $\zeta$. To quantify non-Gaussianity, we compute the cumulants of the one-point probability density function \cite{Bernardeau:2001qr}:
	\begin{equation}
		\label{eq:momenta}
		\kappa_3 = \frac{\langle \zeta^3 \rangle}{\sigma^{3}},\quad \kappa_4=\frac{\langle \zeta^4 \rangle-3\sigma^4}{\sigma ^{4}},\quad \kappa_5=\frac{\langle \zeta^5 \rangle-10 \langle \zeta^3 \rangle\sigma^2}{\sigma ^{5}},
	\end{equation}
	which we normalized by powers of $\sigma^2=\langle \zeta^2 \rangle$ to make them dimensionless.
	In the right panel of \cref{fig:hist_weak} we show the evolution of the cumulants during the simulation. We find that $\kappa_6>\kappa_5>\kappa_4>\kappa_3$ at late times. For illustrative purpose, we avoid showing the evolution of $\kappa_6$, whose final value is $\kappa_6\simeq 10.3$. This result shows that higher-order statistics are essential to characterize non-Gaussianity of $\zeta$. This has important observational consequences, as discussed below.

	
	\section{Strong backreaction}
	We now turn to the case of strong backreaction. We set the gauge coupling to $\alpha/f=25$, 
	so that the imprints of the Chern-Simons coupling on $\zeta$ are unobservable at CMB scales
	 \cite{Anber_2010,Barnaby_2011_Large, Barnaby_2011, Anber_2012}. Later during inflation, however, $\xi$ increases and the universe eventually enters a nonlinear phase.
	
	We start the simulation when $\phi=-5.5$. 
	With this choice, the universe is still in the weak backreaction phase at the beginning of the simulation. Then, after roughly 2 e-folds, the system enters a strong backreaction phase where the bound of \cref{eq:bounds} is violated and \cref{eq:ps_th} gives $\mathcal{P}_\zeta\sim 0.1$, which indicates a breakdown of perturbativity. We show results from a run with $(N,L)=(256,1.5/m)$, but we tested our simulation also with other values of $(N,L)$ to ensure that our results are physical and do not depend on the spatial resolution. Moreover, we ensured the stability of the time integration by checking energy conservation and time-step convergence.
	
	
	In the right panel of \cref{fig:backgroundvalues} we show the evolution of $\xi$ during the simulation. We find the departure from the slow-roll trajectory as an oscillatory behavior in $\xi$. This is intuitive, as one can see from \cref{eq:nonlin} that a strong $F\tilde F$ leads to a depletion of the inflaton velocity; this lowers the value of $\xi$ and reduces the backreaction, bringing the system momentarily back to the slow-roll trajectory. Oscillations of similar period and size were already predicted by previous studies \cite{Cheng:2015oqa,Notari:2016npn,DallAgata:2019yrr,domcke2020resonant}, which explored backreaction effects using semi-analytical tools.
	Another consequence of the backreaction is that, after $6.5$ e-folds of evolution, the background inflaton value is $\phi=-3.02$. This value would be reached after $5.4$ e-folds of evolution if the backreaction were negligible, which means that the backreaction significantly delays the background dynamics. 
	
\begin{figure}
		\centering
		\includegraphics[width=8cm]{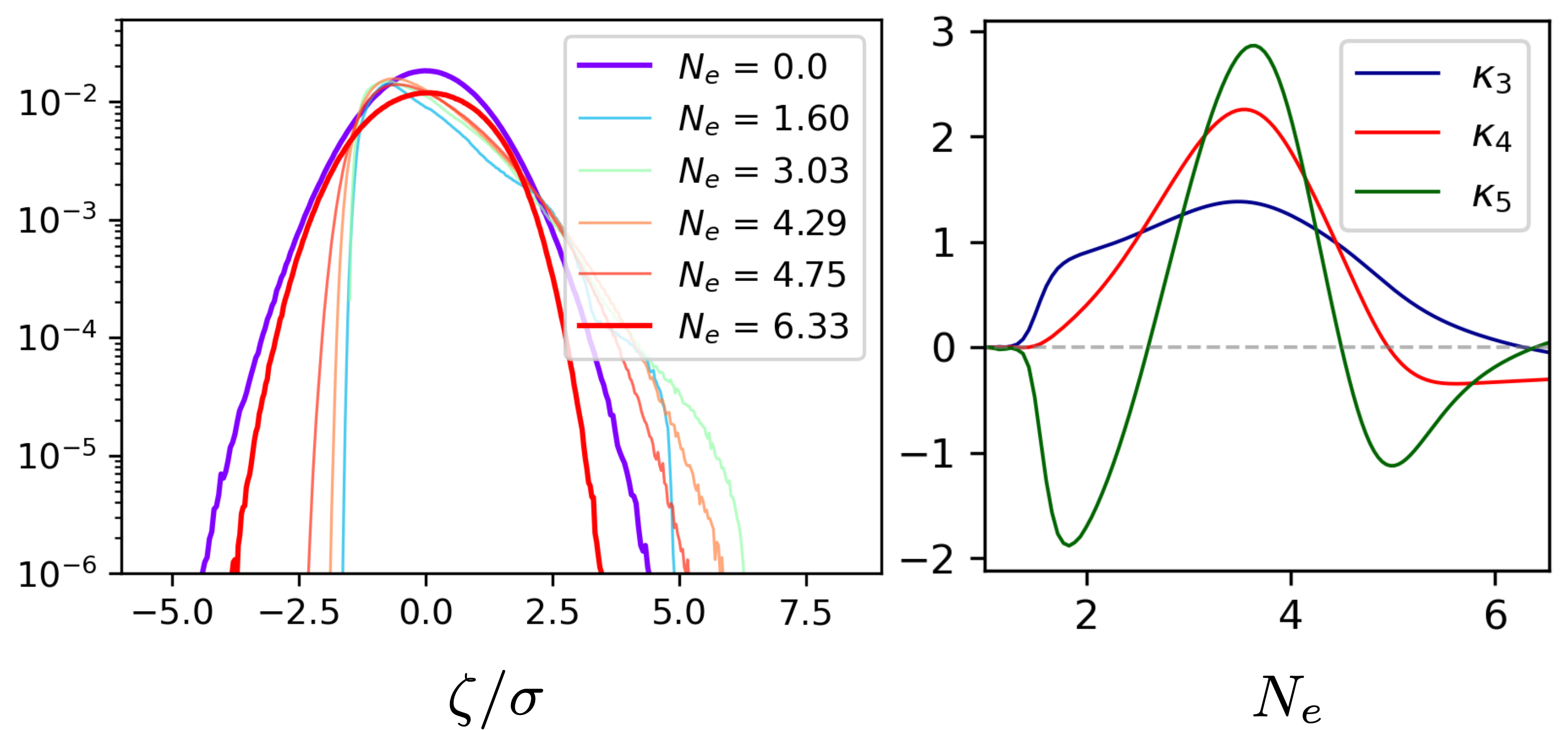}

		\caption{Histograms of $\zeta$ (left) and the cumulants $\kappa_i$ (right) in the case of strong backreaction.}
		\label{fig:hist_strong}
	\end{figure}

	In \cref{fig:hist_strong} we show the histograms of $\zeta$ and the evolution of the cumulants $\kappa_i$. These plots show that the non-Gaussianity of $\zeta$ substantially decreases during the strong backreaction phase. 
	At late times, it is mainly described by a (small) negative $\kappa_4$, while the other cumulants are negligible. Moreover, $\kappa_5$ shows oscillations. The suppression of non-Gaussianity in this regime is a consequence of the central limit theorem, and it is caused by the fact that the number of excited gauge field modes grows with $\xi$. 
	To understand this, we expand the source term $F\tilde F$ in Fourier space as follows:
	\begin{equation}
		\left(	F_{\mu\nu}\tilde F^{\mu\nu}\right)(k)=\sum_{k^{\prime}} F_{\mu\nu}(k^{\prime})\,\,\tilde F^{\mu\nu}(k-k^{\prime}).
	\end{equation}
	This shows that each Fourier mode of $F\tilde F$ is the sum of several non-Gaussian quantities. 
	For $\xi\sim 1$, there are few elements contributing to this sum due to the small number of excited gauge field modes. For $\xi\gg 1$, the number of statistically independent elements in this sum is large, and $F\tilde F$ converges to a Gaussian distribution due to the central limit theorem. Therefore, it sources a Gaussian $\zeta$. 
	This is analogous to what happens when $\phi$ is coupled to fermionic fields \cite{Adshead:2018oaa}.
	
Parametrizing non-Gaussianity in the local type form \cite{Komatsu:2001rj}:
	\begin{equation}
		\label{eq:fnl}
		\zeta(\vec{x})=\zeta_g(\vec{x})+f_{\rm NL}[\zeta_g^2(\vec{x})-\langle\zeta^2_g(\vec{x})\rangle],
	\end{equation}
	where $\zeta_g$ is a Gaussian field and $f_{\rm NL}$ a real number, one can use the linear results of Ref. \cite{Barnaby_2011_Large} to show that $f_{\rm NL}\propto e^{-2\pi\xi}$. This shows that the suppression of non-Gaussianity for large $\xi$ can be guessed from the linear regime of the theory in the $\xi$-constant approximation.
	Although the linear theory is not reliable in this regime, and we find that non-Gaussianity is not of the local type, the simulation confirms this intuition.

\begin{figure}
		\centering
		\includegraphics[width=8cm]{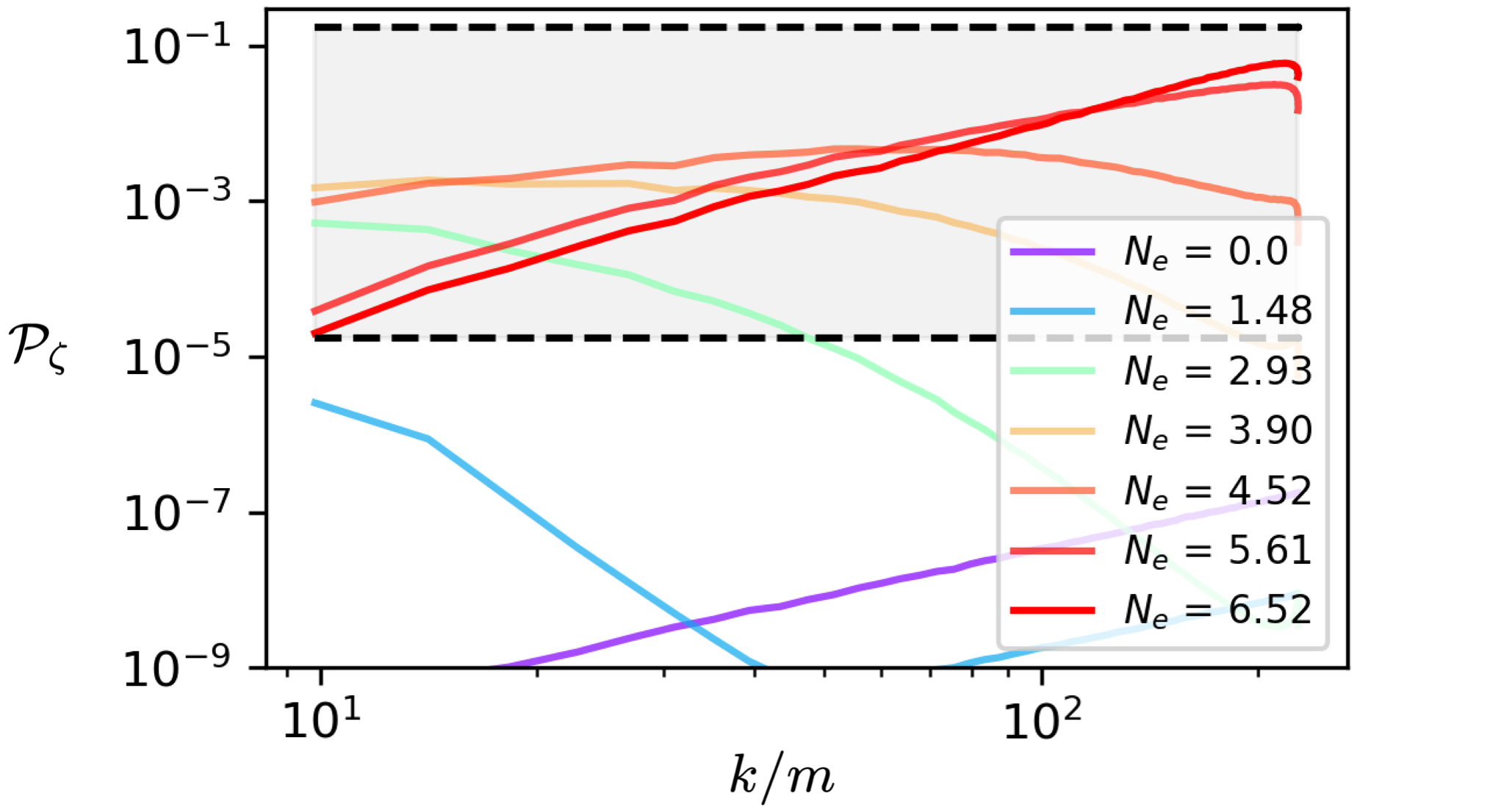}
		
		\caption{Power spectrum 
			from the simulation in the case of strong backreaction.}
		\label{fig:ps_strong}
	\end{figure}

	In \cref{fig:ps_strong} we show the power spectrum from the simulation. Although the analytical estimates are not reliable in this regime, we still compare it with \cref{eq:ps_th} using the initial and final values of $\xi$ from the simulation. 
	As $\kappa_3\ll1$, we are not able to compute the bispectrum because it is below the noise of our bispectrum estimator. 
	
	\section{Discussion}
	 We presented the first lattice simulation of nonlinear perturbations generated during inflation.  For the first time, we obtain the full primordial density fluctuation predicted by inflation.
		We used the simulation to fully characterize the statistics of $\zeta$ from the axion-U(1) model.
		
		 In the weak backreaction regime, power spectrum and bispectrum agree with analytical estimates. Nevertheless, high-order information is crucial in describing the statistics, showing that $n$-point correlators are not efficient to fully characterize this signal. Current large-scale bounds, constraining $\alpha/f\lesssim 32$ for the quadratic potential, are derived using power spectrum and bispectrum of the CMB, ignoring higher-order statistics \cite{Anber_2010,Barnaby_2011_Large, Barnaby_2011, Anber_2012}.
		 
		  In light of our results, we expect the information beyond power spectrum and bispectrum to play a key role. The output of the simulation can be used as truly \textit{ab initio} initial condition for the cosmological simulations of structure formation, allowing to test inflation using the full information contained in the density field. This opens a new possibility for the field of cosmological simulations, that we plan to explore in future work.
	
	
	
	
	
	In the case of strong backreaction, typically relevant for small scales, the system enters an instability phase, characterized by oscillations in the inflation background velocity. In this regime, non-Gaussianity of $\zeta$ is strongly suppressed. This is a consequence of the central limit theorem, and it is caused by the large number of excited gauge field modes contributing to the source term $F\tilde{F}$. 
	We expect this suppression to be a general feature of models where matter fields are coupled linearly to the inflaton $\mathcal{L}\supset \phi f(X)$, with $f(X)$ being a quadratic function of a generic matter field $X$, that could be for example a scalar $X=\psi$ or a gauge field $X=A^{a}_\mu$. If $X$ is copiously produced during inflation via some mechanism, its contribution to the statistics of $\zeta$ is expected to be Gaussian for the same reason. Due to the simplicity of this argument, the same conclusion could hold for more generic nonlinear functions $f(X)$, although this needs to be investigated in future studies.
	
	This result relaxes the bounds from the overclosure of the universe due to overproduction of primordial black holes (PBH), constraining $\alpha/f\lesssim 23$ for the quadratic potential \cite{PhysRevD.87.103506, Garcia-Bellido:2016dkw}. These bounds strongly rely on the assumption that $\zeta$ can be approximated by a (non-Gaussian) $\chi^2$ distribution during the strong backreaction phase \cite{PhysRevD.87.103506, Garcia-Bellido:2016dkw}, which corresponds to the $f_{\rm NL}\gg 1$ limit of \cref{eq:fnl}. 
	We find a nearly Gaussian $\zeta$, requiring a much larger power spectrum at small scales to efficiently produce PBH \cite{PhysRevD.87.103506}. Estimating PBH production requires a more extensive and detailed study of the final e-folds of inflation, for two main reasons. First, as we show in this paper, backreaction significantly delays the end of inflation, making it problematic to identify the range of modes relevant for PBH production. Second, as we show in \cref{fig:hist_strong}, there is still a small remnant non-Gaussianity at the end of the simulation, to which the production of PBH is extremely sensitive.
		
		
	
	
	We conclude that the most stringent bounds on $\alpha/f$ are the ones from the statistics of $\zeta$ at large scales, discussed above. 
		This allows for an inflationary gravitational waves (GW) signal within reach of LISA \cite{LISA1,Bartolo:2016ami}, advanced LIGO \cite{LIGOScientific:2016fpe} and PTA-SKA \cite{1990ApJ...361..300F,5136190,Kramer:2004hd} experiments. Indeed, the gauge field acts as a source for GW \cite{Anber_2010,Barnaby_2011_Large, Barnaby_2011, Anber_2012}, and the signal can be above the projected sensitivity of all these experiments in the parameter range compatible with current CMB constraints \cite{Garcia-Bellido:2016dkw}.

	Note that, both for simplicity and to better compare with previous studies, we considered a quadratic potential for the inflaton $V(\phi)=\frac{1}{2}m^2\phi^2$, which is disfavored by the latest Planck-BICEP/Keck results \cite{Planck:2018jri,BICEP:2021xfz}. The particular choice of slow-roll potential, however, only affects the quantitative bounds on $\alpha/f$ given above. Our findings about the statistics of $\zeta$, that are the main original result of this work, do not depend on this choice.

	The lattice simulation presented in this paper allowed to reveal unknown aspects of the axion-U(1) system that are beyond the regime of validity of perturbation theory.
	At the same time, it allowed to compute inflationary observables within this model with a precision that exceeds state-of-the-art analytical and semi-analytical computations, both in the weak and strong backreaction regimes. 
There several other cases where a lattice simulation could be crucial, like models involving non-abelian SU(2) gauge fields \cite{Maleknejad_2011,maleknejad2013gaugeflation,Adshead_2012,Adshead_2013,maleknejad2021su2r}, or scalar fields models with a strong turn in field space \cite{Fumagalli:2020nvq}. 	Our work shows that lattice simulations could be an essential tool to understand the predictions of these models, and more generically inflationary scenarios characterized by nonlinear physics.

	\begin{acknowledgments}
		This work is supported in part by the Excellence Cluster ORIGINS which is funded by the Deutsche Forschungsgemeinschaft (DFG, German Research Foundation) under Germany's Excellence Strategy - EXC-2094 - 390783311 (AC, EK, JW), and JSPS KAKENHI Grant Number JP20H05859 (EK). The Kavli IPMU is supported by World Premier International Research Center Initiative (WPI), MEXT, Japan. The work of KL was supported in part by the US Department of Energy through grant
		DE-SC0015655.
	\end{acknowledgments}

	\bibliographystyle{apsrev4-2}
	\bibliography{U1Sim_PRL}

\end{document}